\documentclass[%
 reprint,
 superscriptaddress,
 groupedaddress,
 nofootinbib,
 longbibliography,
 amsmath,amssymb,
 aps,
 prd,
]{revtex4-1}

\usepackage{xcolor}
\usepackage{graphicx}
\usepackage{dcolumn}
\usepackage{bm}
\usepackage{hyperref} %black hyperref
\usepackage{url}

\newcommand{\beq}{\begin{equation}}
\newcommand{\eeq}{\end{equation}}

\begin{document}

\begin{flushright}
IFUM-1090-FT
\end{flushright}

\title{\Large {Binary black hole system at equilibrium}}

\author{Marco Astorino}
 \email{marco.astorino@gmail.com}
 \affiliation{%
 Istituto Nazionale di Fisica Nucleare (INFN), Sezione di Milano \\
 Via Celoria 16, I-20133 Milano, Italy
}%
\author{Adriano Vigan\`o}%
 \email{adriano.vigano@unimi.it}
 \affiliation{%
 Universit\`a degli Studi di Milano \\
 %Via Celoria 16, I-20133 Milano, Italy
}%
 \affiliation{%
 Istituto Nazionale di Fisica Nucleare (INFN), Sezione di Milano \\
 Via Celoria 16, I-20133 Milano, Italy
}%

\begin{abstract}

An exact and analytical solution of four dimensional vacuum General Relativity representing a system of two static black holes at equilibrium is presented.
The metric is completely regular outside the event horizons, both from curvature and conical singularities.
The balance between the two Schwarzschild sources is granted by an external gravitational field, without the need of extra matter fields besides gravity, nor strings or struts.
The geometry of the solution is analysed.
The Smarr law, the first and the second law of black hole thermodynamics are discussed.

\end{abstract}

\maketitle

\section{Introduction}

Black holes are fundamental objects of our Universe.
Nevertheless there are scarce exact and analytical solutions that model black holes within the context of General Relativity, which represents the standard framework of gravitational interactions.
The requirement of asymptotic flatness further restricts the number of available solutions, basically due to the existence of uniqueness theorems~\cite{Chrusciel:2012jk}.

The recent remarkable observational results from gravitational waves detection reveal that the major source of  gravitational waves is provided by the interaction between massive black holes~\cite{Abbott:2016blz}.
Unfortunately, there are no exact analytical solutions describing well-founded multi-source black holes in pure General Relativity.
The few attempts built so far need singular matter to sustain the gravitational attraction between the sources, as can be seen from double Kerr attempts, see~\cite{Kramer,Dietz}.
This matter is often interpreted as struts or cosmological strings, however these objects not only have stability issues and no experimental plausibility at the moment; they also rise some fundamental theoretical issues related to the singular behaviour of the spacetime, which therefore can hardly be thought as a proper manifold.
The $\delta$-like stress-energy tensor associated to the strut, located between the black holes, violates all the reasonable energy conditions because, in order to prevent the gravitational sources from the collapse, it has to mimic a repulsive gravitational force generated by anti-gravitational matter.
Alternatively, the black holes can be pulled by axial strings of proper matter, i.e.~of positive energy density, which however are divergent quantities that extend to infinity.
Both scenarios are clearly out of current empirical experience.

On the other hand, in the presence of electromagnetic fields, it is possible to gain the equilibrium between two black holes thanks to the electromagnetic repulsive force that balances the gravitational attraction.
In fact some solutions of these kind are known, basically the Majumdar--Papapetrou black holes~\cite{Majumdar:1947eu,Papapetrou,Hartle:1972ya} and the Emparan dihole~\cite{Emparan:1999au}.
While these solutions are regular outside the event horizons of the sources, they need to be composed of charged matter, which is improbable from the astrophysical point of view.
Furthermore they are extremal and their physical parameters are very fine tuned to maintain the equilibrium, therefore they are likely to be unstable under perturbations.

The purpose of this Letter is to explore the possibility to build analytical metrics modelling two black hole sources at equilibrium, aiming at the most phenomenologically realistic scenario as possible.
For this reason, we will discard any source of singularities outside the event horizons, not only Dirac or cosmic strings, but also Misner strings.
In fact, the gravimagnetic parameter (NUT) is known to bring in physical issues, e.g.~closed timelike curves.

The strategy of our approach is based on an Ernst's insight, who managed to regularise the nodal singularity of the accelerating Schwarzschild black hole in~\cite{ernst}.
Similarly we examine the possibility of balancing the gravitational attraction of the two (or more) sources by an external gravitational field, which serves as a setting for embedding our multi-source solution.
Thus the spacetime asymptotic will not be restricted to the flat one;
instead we prefer to consider an axisymmetric multi-polar expansion~\cite{TOMIMATSU1984374,breton-manko} able to model a generic gravitational background~\footnote{An interior multi-polar deformation of the sources can be achieved as well~\cite{erez-rosen}, but it seems that these poles bring curvature singularities not covered by an event horizon.}.
The multipolar expansion is the key ingredient that allows to circumvent the no-go theorem about the non-existence of static~\cite{Beig_2009} configurations of many-body systems, with a suitable separation condition.

These kind of solutions are know in the literature, in the single source case, as distorted black holes:
they have been introduced by Doroshkevich, Zel'dovich and Novikov~\cite{NovikovZeldovich}, and lately Chandrasekhar~\cite{Chandrasekhar:1985kt} studied their equilibrium conditions.
Geroch and Hartle~\cite{Geroch:1982bv} enlightened then some of their general properties.
Distorted black holes are especially valuable when the black hole is not isolated but embedded in a strong field regime, such as in the center of galaxies, where approximate, perturbative or numerical methods loose accuracy.
Distorted metrics are generally considered local in the sense that are mostly meaningful in a finite region surrounding the black holes.
The distortion is due to external sources, which are thought to be at large distances from the black holes, and in fact they do not contribute with a stress-energy tensor~\footnote{The situation is analogous to the Melvin magnetic universe, which remains an electrovacuum solution even though the sources of the magnetic field should somehow be present at infinity.}.
A full solution should also include these sources, as done up to the $6^{th}$ order in the multipole expansion in~\cite{deCastro:2011zz} for a thin disk of matter.
In~\cite{Astorino:2021rdg} it is shown that the inclusion of the acceleration parameter interposes a Rindler horizon between the black sources and the spatial infinity.

While the number of gravitational multipoles of the background can in principle be left arbitrary~\cite{Astorino:2021boj}, in this Letter we focus only on the dipole and quadrupole terms since it represents the simplest setting in which the black holes distinctive parameters remain unconstrained.
We stress that our regularisation method relies only on the presence of (at least) two multipole momenta, therefore it can be achieved for any couple on non-null poles.

In practice we superimpose a double black hole solution to the external field background by means of the Belinski--Zakharov ``inverse scattering method''~\cite{Belinsky:1971nt,Belinsky:1979mh,Belinski:2001ph}.
The details of the inverse scattering construction in the context of external gravitational fields are left for a future work~\cite{Astorino:2021boj}.

\section{Metric and regularisation}

Consider the static and axisymmetric metric, solution of the Einstein equations in vacuum $R_{\mu\nu}=0$, in cylindrical Weyl coordinates $(t,\rho,z,\phi)$
\beq
\label{metric}
\begin{split}
{ds}^2 & = - V(\rho,z) {dt}^2 + \rho^2 V^{-1}(\rho,z) {d\phi}^2 \\
& \quad + f(\rho,z) \bigl({d\rho}^2 + {dz}^2\bigr) ,
\end{split}
\eeq
where
\begin{widetext}
\begin{align}
V & = \frac{\mu_1\mu_3}{\mu_2\mu_4}
\exp\biggl[2b_1z + 2b_2\biggl(z^2 - \frac{\rho^2}{2}\biggr)\biggr] , \\
\begin{split}
f & = 16C_f \frac{\mu_1^3\mu_2^5\mu_3^3\mu_4^5}{W_{11}W_{22}W_{33}W_{44}W_{13}^2W_{24}^2Y_{12}Y_{14}Y_{23}Y_{34}}
\exp\biggl[-b_1^2\rho^2 + \frac{b_2^2}{2} \bigl(\rho^2 - 8z^2\bigr)\rho^2 - 4b_1b_2 z \rho^2 \\
& \quad + 2 b_1 (-z + \mu_1 - \mu_2 + \mu_3 - \mu_4 )
+ b_2 \bigl(-2z^2 + \rho^2 + 4z (\mu_1 - \mu_2) + \mu_1^2 - \mu_2^2
+ (\mu_3 - \mu_4) (4z + \mu_3 + \mu_4) \bigr)\biggr],
\end{split}
\end{align}
\end{widetext}
and
$W_{ij}=\rho^2+\mu_i\mu_j$,
$Y_{ij}=(\mu_i-\mu_j)^2$.
The functions
$\mu_i=\sqrt{\rho^2+(z-w_i)^2}-(z-w_i)$
contain the constants $w_i$, that are chosen with ordering $w_1<w_2<w_3<w_4$ and with values
\begin{subequations}
\begin{align}
w_1 & = z_1 - m_1 , \qquad
w_2 = z_1 + m_1 , \\
w_3 & = z_2 - m_2 , \qquad
w_4 = z_2 + m_2 .
\end{align}
\end{subequations}
The parameters $m_i$ and $z_i$ refer to the mass and to the position of the $i$-th black hole, respectively, while $b_1$ and $b_2$ are the dipole and quadrupole momenta of the external gravitational field polar expansion.
The constant $C_f$ is a gauge parameter.

The metric~\eqref{metric} represents two Schwarzschild black holes immersed in an external back-reacting gravitational field~\footnote{A Mathematica notebook containing this metric can be found at~\href{https://sites.google.com/site/marcoastorino/papers/2104-07686}{https://sites.google.com/site/marcoastorino/papers/2104-07686} and as an ancillary file on the arXiv webpage.}.
The two event horizons extend in the regions $w_1<z<w_2$ and $w_3<z<w_4$ when $\rho=0$,
while the curvature singularities are covered by the horizons.
This can be  understood thanks to a criterion given in~\cite{Harmark:2004rm}:
the kernel of the ($t,\phi$) part of the metric can be 2-dimensional only inside the horizon.
By removing the external field ($b_1=b_2=0$), one straightforwardly recovers the double-Schwarzschild solution (also known as Bach--Weyl metric)~\cite{Bach1922,Israel1964,Costa:2000kf}.

A physical solution should be regular on the $z$-axis, i.e.~it must not be affected by conical singularities.
These angular defects are present when the ratio between the length and the radius of small circles around the $z$-axis is different from $2\pi$.
%which corresponds to a globally non-flat geometry.
A small circle around the $z$-axis has radius $R=\sqrt{g_{zz}}\rho$ and length $L=2\pi\sqrt{g_{\phi\phi}}$ (cf.~\cite{Alekseev:2019kcf}),
then the regularity condition corresponds to
$L/(2\pi R)\to 1$ as $\rho\to 0$.
It is easy to prove that, for the diagonal metric~\eqref{metric}, the above condition is equivalent to $fV\to 1$ as $\rho\to 0$.
The latter must be satisfied everywhere, outside the black holes, to avoid conical singularities.
We thus require
\beq
\label{constraint}
fV = 1 ,
\eeq
over the intervals $-\infty<z<w_1$, $w_2<z<w_3$ and $w_4<z<+\infty$ as $\rho\to0$.
Assuming a $2\pi$-periodic azimuthal angle $\phi$, we have to fix three out of seven parameters in order to satisfy~\eqref{constraint}.
\begin{figure*}
\centering
\includegraphics[scale=0.45]{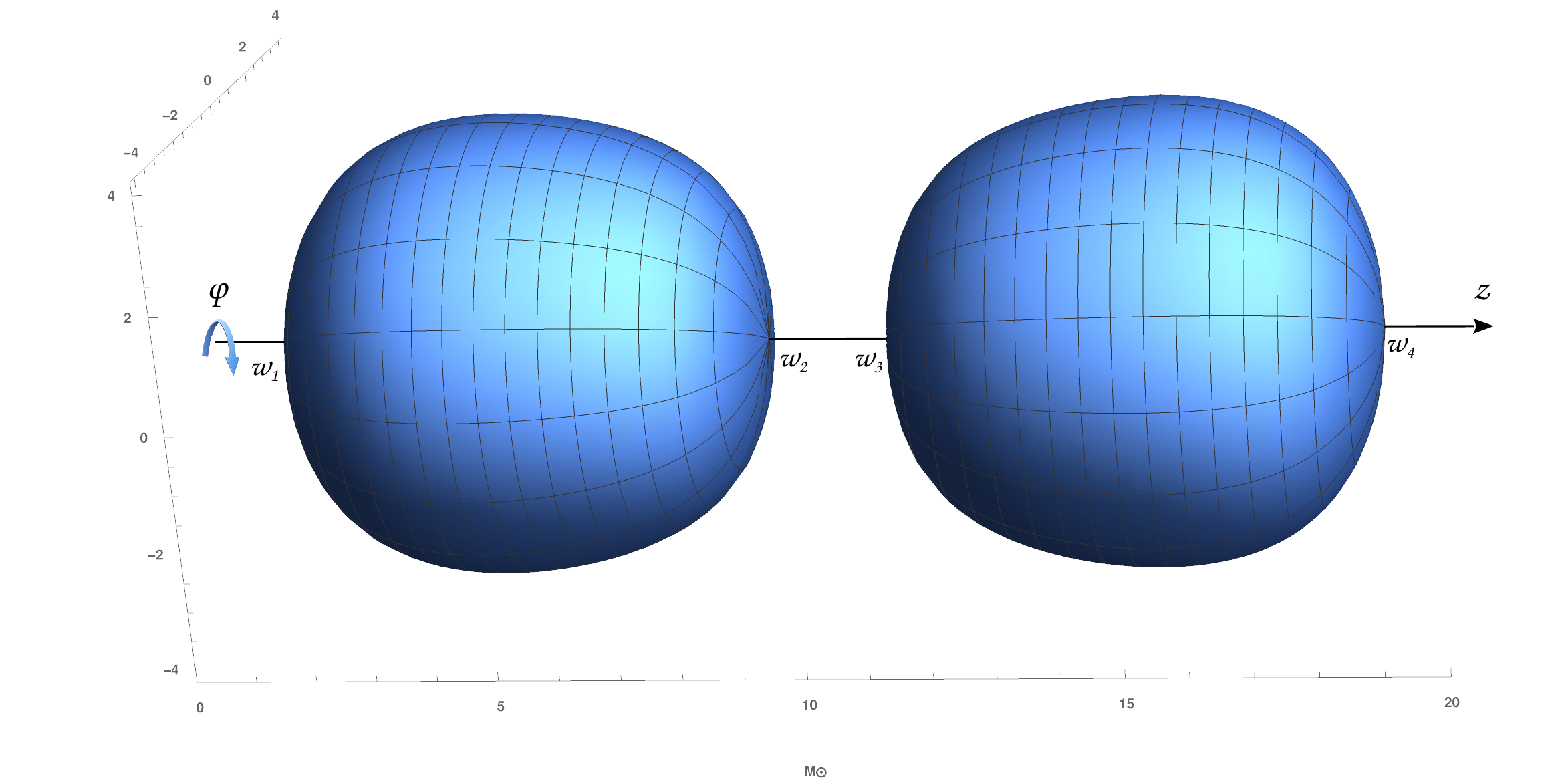}
\caption{\small Embedding diagram in $\mathbb{E}^3$ of the surfaces of two black hole event horizons for the parametric values $z_1=5$, $z_2=15$, $m_1=4$, $m_2=4$, in Solar mass units $M_\odot$. This picture shows the deformation of the horizons due to both the external gravitational field and the mutual interaction between the sources. The horizon surface is smooth because of the absence of any conical singularities.}
\label{fig:embedding}
\end{figure*}
%In the different regions of spacetime, outside the horizons, we find
%\begin{widetext}
%\begin{subequations}
%\begin{align}
%P_{z<w_1} & =
%\biggl[\frac{C_f e^{-4 (b_1(m_1+m_2) + 2b_2 (m_1z_1+m_2z_2))}}{16m_1m_2 (m_1+m_2+z_1-z_2)(m_1+m_2-z_1+z_2)}\biggr]^2 \\
%P_{w_2<z<w_3} & =
%\biggl[\frac{C_f e^{-4 m_2(b_1+2b_2z_2)}}{16m_1m_2 (m_1-m_2+z_1-z_2)(m_1-m_2-z_1+z_2)}\biggr]^2 \\
%P_{z>w_4} & = \biggl[\frac{C_f}{16m_1m_2 (m_1+m_2+z_1-z_2)(m_1+m_2-z_1+z_2)}\biggr]^2 .
%\end{align}
%\end{subequations}
%\end{widetext}
We choose the gauge parameter $C_f$ and the external field parameters $b_1$, $b_2$ as follows:
\begin{widetext}
\begin{align}
C_f & = 256 m_1^2 m_2^2 (m_1+m_2+z_1-z_2)^2(m_1+m_2-z_1+z_2)^2 , \\
\label{b1}
b_1 & = -\frac{(m_1z_1+m_2z_2)}{4m_1m_2(z_1-z_2)}
\log\biggl[\frac{(m_1-m_2+z_1-z_2)(m_1-m_2-z_1+z_2)}{(m_1+m_2+z_1-z_2)(m_1+m_2-z_1+z_2)}\biggr] , \\
\label{b2}
b_2 & = \frac{(m_1+m_2)}{8m_1m_2(z_1-z_2)}
\log\biggl[\frac{(m_1-m_2+z_1-z_2)(m_1-m_2-z_1+z_2)}{(m_1+m_2+z_1-z_2)(m_1+m_2-z_1+z_2)}\biggr] .
\end{align}
\end{widetext}
It is understood that $C_f$, $b_1$, $b_2$ will assume the above values, from now on.
Note that the physical parameters characterising the black holes were left unconstrained:
this means that the physical properties of the binary black hole system are completely generic.
This fact guarantees a wide flexibility in a possible phenomenological scenario~\footnote{In fact the presented solution is not the minimal one which can be regularised.
It is sufficient to consider only one multipole term for the external  gravitational field to remove all the singularities from the double black hole configuration, at the price of fixing the position or the mass of a black hole.}.
Of course one could alternatively keep the $b_i$ generic, to model an arbitrary external gravitational background:
in that case the intrinsic black holes parameters would adjust to fit the given background.

One can infer the order of magnitude of the multipole momenta~\eqref{b1},~\eqref{b2} for an experimental setting.
We make an explicit example by considering the data for the black holes masses and separation provided by the GW150914 event~\cite{Abbott:2016blz}:
we take $m_1=29 M_\odot$, $m_2=36 M_\odot$, $|z_1-z_2|=520 M_\odot$, and we find
$b_1\sim 10^{-4}/M_\odot$ and
$b_2\sim -10^{-7}/M_\odot^2$.
These are the values that ensure the equilibrium, therefore in the case of a merging process they represent only an upper bound.
The relevance of the contributions of $b_1$ and $b_2$ in the context of the Geroch--Hansen multipoles~\cite{Hansen:1974zz} will be discussed in~\cite{Astorino:2021boj}.

%One might wonder if the dipole-quadrupole configuration described by~\eqref{b1},~\eqref{b2} can be sustained by physical matter, i.e.~by matter satisfying the standard energy conditions~\cite{Hawking:1973uf}.
%According to~\cite{Geroch:1982bv}, the sources of the external field satisfy the strong energy condition if $b_2<0$:
%this condition is always satisfied for values of the parameters that mantain the horizon areas and the temperatures positive (see below).
%We thus conclude that the field supporting the double black hole system can be generated by regular matter.

We notice that it is possible, within our scheme, to build an array of collinear black holes immersed in an external gravitational field with complete multipolar expansion~\cite{Astorino:2021boj}, thus generalising the Israel--Kahn solution~\cite{Israel1964}.
In that case, the regularisation would require the complete multipole expansion~\footnote{Actually an arbitrary multipole expansion can be useful to model any generic stationary and axisymmetric external gravitational field.}
of the external field to fix the conical singularities, without constraining the physical parameters of the black holes, such as their masses or positions on the $z$-axis.

To ensure phenomenological significance of the solution~\eqref{metric}, it is essential that the separation of the two sources is finite.
The proper distance between the two black hole horizons for $\rho=0$ and fixed $\phi$ is given by
\beq
\begin{split}
\ell & = \int_{w_2}^{w_3} dz \, \sqrt{g_{zz}} \big|_{\rho=0}
\propto
\int_{w_2}^{w_3} dz \, \sqrt{\frac{(w_4-z)(z-w_1)}{(w_3-z)(z-w_2)}} \\
&\quad \times e^{2b_1(2w_3-2w_4-z) + 2b_2(2w_3^2-2w_4^2-z^2)}.
\end{split}
\eeq
It can be shown that the integral converges, and hence the equilibrium can be achieved for finite proper distance.

\section{Near-horizon limit}

We consider the near-horizon limit of the metric~\eqref{metric}, in order to show that it contains two distorted Schwarzschild black holes.
We zoom in to the first black hole horizon by performing the change of coordinates
\beq
\rho = \sqrt{r(r-2m_1)}\sin\theta , \quad
z = z_1+(r-m_1)\cos\theta ,
\eeq
and by taking the limit $r\to 2m_1$, by which the metric~\eqref{metric} boils down to
\beq
\label{nh}
\begin{split}
{ds}^2 & \simeq h(\theta) \biggl[ -\biggl(1-\frac{2m_1}{r}\biggr) e^{F_1(\theta)} {dt}^2
+ \frac{D^2 e^{F_2(\theta)}}{1-2m_1/r} {dr}^2 \biggr] \\
&\quad + (2m_1)^2 \biggl[ D^2 h(\theta) e^{F_2(\theta)} {d\theta}^2
+ \frac{\sin^2\theta}{h(\theta)} e^{-F_1(\theta)} {d\phi}^2 \biggr] ,
\end{split}
\eeq
where
\begin{align}
h(\theta) & = \frac{m_1\cos\theta+m_2+z_1-z_2}{m_1\cos\theta-m_2+z_1-z_2} , \\
F_1(\theta) & = 2 [b_1 + b_2(z_1+m_1\cos\theta)] (z_1 + m_1\cos\theta) , \\
\begin{split}
F_2(\theta) & = 2 b_1 (m_1\cos\theta - 2m_1 - 4m_2 - z_1) \\
&\quad + 2 b_2 (m_1^2\cos^2\theta + 2m_1z_1\cos\theta - 2m_1^2 \\
&\quad - z_1^2 - 4m_1z_1 - 8m_2z_2) ,
\end{split}
\end{align}
and
\beq
D = \frac{m_1+m_2-z_1+z_2}{m_1-m_2-z_1+z_2} .
\eeq
One clearly recognises in~\eqref{nh} the structure of a distorted Schwarzschild black hole~\cite{Geroch:1982bv}.
Actually the first black hole horizon is deformed by the presence of both the external field and the second black hole.
Indeed, when the external field and the second black hole vanish, one recovers the standard Schwarzschild metric.
Obviously, a similar description holds for the second black hole as well.
A pictorial representation of the deformation that the two horizons undergo is given in the embedding diagram of Fig.~\ref{fig:embedding}.

Another element about the geometry of the black hole horizon can be clarified by computing the length of the equatorial and polar circles in the near-horizon geometry~\eqref{nh}.
The equator ($\theta=\pi/2$) length is given by
\beq
\label{equator}
\begin{split}
L_\text{equator} & = 2m_1 \int_0^{2\pi} \frac{e^{-F_1(\pi/2)/2}}{\sqrt{h(\pi/2)}} d\phi\\
& = 4\pi m_1 \sqrt{\frac{z_1-z_2-m_2}{z_1-z_2+m_2}}
e^{-2(b_1+b_2z_1)z_1} ,
\end{split}
\eeq
while the polar length is
\beq
\label{polar}
L_\text{polar} = 4m_1 D \int_0^\pi \sqrt{h(\theta)} e^{F_2(\theta)/2} d\theta .
\eeq
It is not possible to analytically perform the latter integral, nevertheless we can consider a numerical comparison between~\eqref{equator},~\eqref{polar} and $L_\text{Schwarzschild}=4\pi m_1$.
The result is
\begin{subequations}
\label{comparison}
\begin{align}
L_\text{equator} & > L_\text{Schwarzschild} , \\
L_\text{polar} & > L_\text{Schwarzschild} .
\end{align}
\end{subequations}
Thus, not only a deformation along the $z$-axis occurs, but there is also an enlargement of both the equatorial and the polar circle with respect to the Schwarzschild one.
This is consistent with the behaviour of the black hole temperature, as we will see below.

\section{Smarr law and thermodynamics}

Going back to the exact metric~\eqref{metric}, we now compute some physical quantities for that spacetime, in order to discuss the thermodynamics of the system.

The mass of the two black holes is found by means of the Komar--Tomimatsu integral~\cite{Komar,Tomimatsu:1984pw}.
The formula for the conserved mass, in the case of the static spacetime~\eqref{metric}, takes the form
\beq
M = \alpha \int_{w_{2i-1}}^{w_{2i}} dz\, \rho\, g_{tt}^{-1}\, \partial_\rho g_{tt} \big|_{\rho=0} ,
\eeq
where $\alpha$ is a constant that takes into account the normalisation of the timelike Killing vector $\xi\equiv\alpha\partial_t$, which generates the stationary symmetry.
It is well known that, in the absence of asymptotic flatness, $\alpha$ is not necessarily equal to one, as happens for black holes in AdS~\cite{Henneaux:1985tv} or Melvin~\cite{Ashtekar:2000hw,Astorino:2016hls} backgrounds.
The integration is performed over the intervals $w_1<z<w_2$ and $w_3<z<w_4$, which correspond to the black holes, and the result is
\beq
\label{mass}
M_1 = \alpha m_1 , \qquad
M_2 = \alpha m_2 .
\eeq
The horizons areas are found integrating over the horizon surfaces, i.e.
\beq
%\begin{split}
\mathcal{A} = \int_{0}^{2\pi} d\phi
\int_{w_{2i-1}}^{w_{2i}} dz\, \sqrt{g_{zz}g_{\phi\phi}} \Big|_{\rho=0} ,
%& = \int_{0}^{2\pi} d\phi
%\int_{w_i}^{w_{i+1}} dz\, \rho\sqrt{f V^{-1}} \Big|_{\rho=0} ,
%\end{split}
\eeq
thus
\begin{align}
\begin{split}
\mathcal{A}_1 & =
16\pi m_1^2 \frac{m_1+m_2-z_1+z_2}{m_1-m_2-z_1+z_2} \\
&\quad \times e^{-2b_1 (2m_2+m_1+z_1)-2b_2((m_1+z_1)^2+4m_2z_2)} ,
\end{split} \\
\begin{split}
\mathcal{A}_2 & =
16\pi m_2^2 \frac{m_1+m_2-z_1+z_2}{m_2-m_1-z_1+z_2} \\
&\quad \times e^{-2b_1(m_2+z_2)-2b_2(m_2+z_2)^2} .
\end{split}
\end{align}
The temperature is obtained from the surface gravity as $T=\kappa/(2\pi)$.
Recalling that $\kappa^2=-\frac{1}{2} (\nabla_\mu \xi_\nu)^2$,
the metric~\eqref{metric} gives rise to
\beq
\kappa^2 = -\frac{\alpha^2}{4} \frac{(\partial_z V)^2+(\partial_\rho V)^2}{fV} \bigg|_{\rho=0} .
\eeq
%that has to be evaluated over $w_i<z<w_{i+1}$.
The temperatures are then
\begin{align}
\begin{split}
T_1 & =
\frac{\alpha}{8\pi m_1} \frac{m_1-m_2-z_1+z_2}{m_1+m_2-z_1+z_2} \\
&\quad \times e^{2b_1 (2m_2+m_1+z_1)+2b_2((m_1+z_1)^2+4m_2z_2)} ,
\end{split} \\
\begin{split}
T_2 & =
\frac{\alpha}{8\pi m_2} \frac{m_2-m_1-z_1+z_2}{m_1+m_2-z_1+z_2} \\
&\quad \times e^{2b_1(m_2+z_2)+2b_2(m_2+z_2)^2} .
\end{split}
\end{align}
One can verify that the same results are found via the Euclidean method~\cite{Hartle:1976tp}.
We notice that the presence of the external field lowers the black holes temperature, with respect to the Schwarzschild one, and hence the surface gravity.
A lower surface gravity means a lower gravitational ``pressure'' on the horizon, which then can swell up.
This feature is in agreement with~\eqref{comparison} and with the related observations.
Moreover, it explains how the external gravitational field acts, providing an external pressure in the region of the holes, to sustain the mutual gravitational collapse of the binary system. 

Defining the entropy as $S=\mathcal{A}/4$, the above quantities satisfy the Smarr law~\cite{Smarr:1972kt} both for the individual black holes $M_i=2 T_i S_i$ ($i=1,2$)
%\beq
%M_1 = 2 T_1 S_1 , \qquad
%M_2 = 2 T_2 S_2 .
%\eeq
and for the double configuration
\beq
M_1 + M_2 = 2 T_1 S_1 + 2 T_2 S_2 .
\eeq
This result holds regardless of the value of the constant $\alpha$.
Nevertheless, a choice for $\alpha$ must be done in order to study the thermodynamics of the system.

We are interested in the first law of thermodynamics from a local point of view:
the involved quantities are evaluated on the horizons, therefore the sources at infinity (which generate the external field) are not accessible to local observers near the black holes.
Hence we will discard work terms, in the first law, due to the variation of the parameters $b_i$~\cite{Geroch:1982bv}.

We consider the system at thermal equilibrium from now on, i.e.~$T_1=T_2\equiv\bar{T}$.
This condition is satisfied by imposing $m_1=m_2$. We furthermore choose
\beq
\begin{split}
\alpha & = \sqrt{\frac{z_1-z_2-2m_1}{z_1-z_2}} \\
& \quad\times e^{-(b_1+b_2(m_1+z_2))(m_1+z_2)} ,
\end{split}
\eeq
in order to fulfill a Christodoulou--Ruffini mass formula~\cite{Christodoulou:1972kt}, as it happens for regular metrics in which the asymptotic symmetry is different from the flat one~\cite{Astorino:2016hls,Caldarelli:1999xj,Astorino:2016ybm}.
For black hole configurations endowed with $N$ disconnected horizons the best proposal is an addictive generalisation such that
\beq
\label{cristo-N}
\sum_{i=1}^{N} M_i =
\sum_{i=1}^{N} \sqrt{\frac{\mathcal{A}_i}{16 \pi}} \, .   
\eeq
It is worth noting that $m_1=m_2$ is not the only possibility for thermal equilibrium, but it is clearly the simplest one.
Moreover, these choice guarantees the integrability of the masses~\eqref{mass}.

Defining the total mass
$\bar{M}\equiv M_1+M_2$ and the total entropy
$\bar{S}\equiv S_1 + S_2$,
the first law of thermodynamics
\beq
\label{firstlaw}
\delta\bar{M} = \bar{T} \delta\bar{S} ,
\eeq
is verified.
We notice that the variation in~\eqref{firstlaw} is taken with respect to the free parameters $m_1$, $z_1$, $z_2$, in which $\alpha$ depends:
%$\delta=\frac{\partial}{\partial m_1}\delta m_1+\frac{\partial}{\partial z_1}\delta z_1+\frac{\partial}{\partial z_2}\delta z_2$.
thus the presence of $\alpha$ is crucial to the first law.

We now turn to the verification of the second law of thermodynamics. At this scope we consider a process in which the initial state is described by two black holes at finite distance with total mass $\bar{M}$ and entropy $\bar{S}$,
while the final state is modeled by a single black hole of mass
\beq
M_0 = m_0 e^{-(b_1+(m_0+z_0)b_2)(m_0+z_0)} ,
\eeq
and entropy
\beq
S_0 = 4\pi m_0^2
e^{2(b_1+(m_0+z_0)b_2)(m_0+z_0)} .
\eeq
Here $m_0$ and $z_0$ are the mass parameter and the position of the single black hole, respectively. $M_0$ and $S_0$ are computed from metric~\eqref{metric} with a single black hole.

Note that, contrary to~\cite{TOMIMATSU1984374,breton-manko}, we do not assume $z_0=0$ \emph{a priori}.
The presence of the external gravitational field breaks the translation invariance along the $z$-axis, since the field acts differently on different points of the axis.
\begin{figure}
\centering
\includegraphics[scale=0.27]{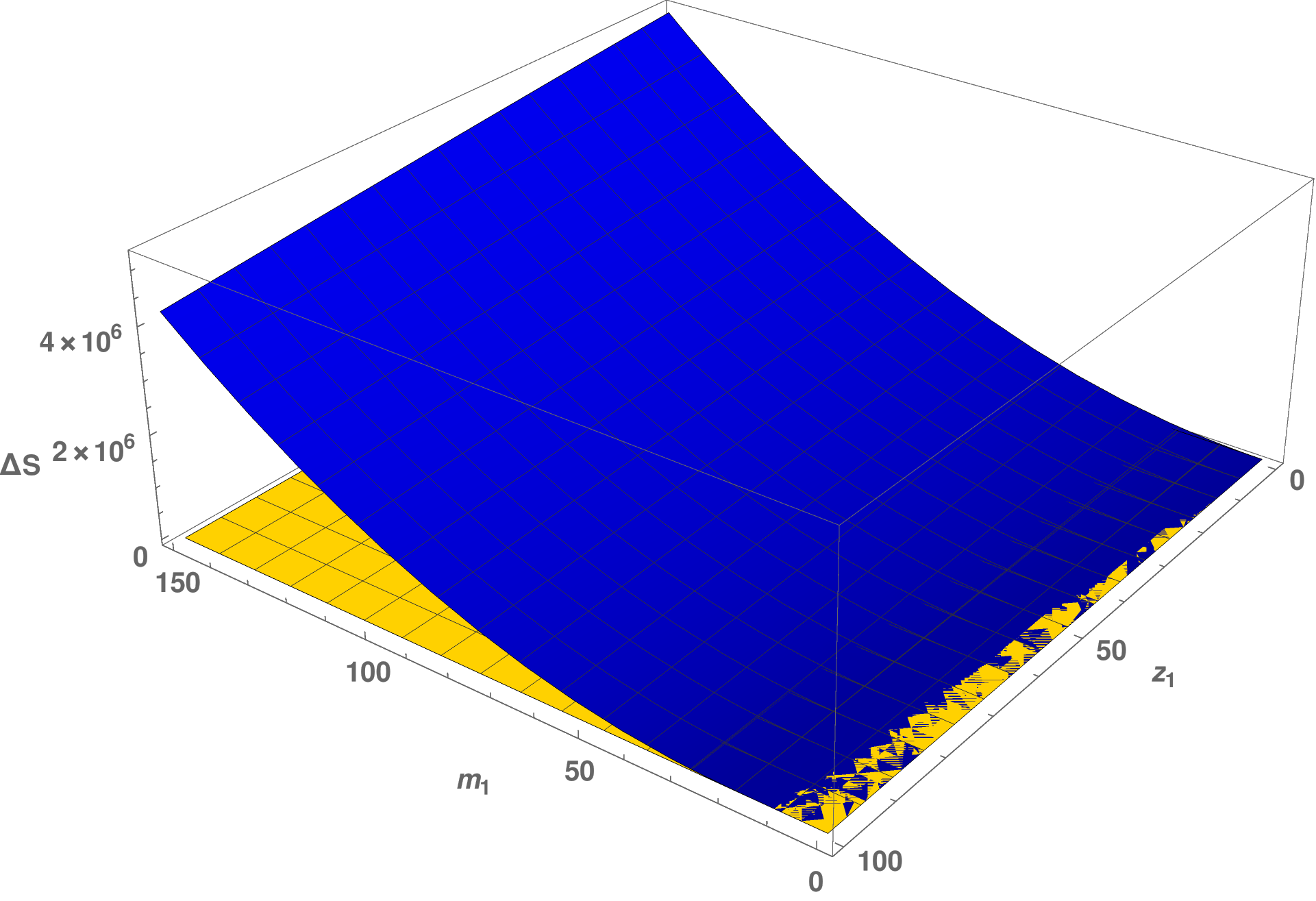}
\caption{\small The blue surface represents the entropy variation $\Delta S$, while the yellow surface is the 0 plane. The parameters vary over the values $m_1\in[0,150]$ and $z_1\in[0,100]$. Within the whole parametric range the single black hole state results more entropic than the disjoint pair.}
\label{fig:entropy}
\end{figure}
This fact is relevant, because it allows to regularise the single black hole metric with $b_1\neq0$.
In the single black hole case, the regularisation condition is
$b_1 = -2 b_2 z_0$.

In order to have a meaningful comparison we need to compare the two states at equal energy~\footnote{We are assuming that there is no energy loss due to emitted radiation.} and also their background must coincide.
Being $b_2$ unconstrained in the single black hole configuration, we just have to equate the values for $b_1$, which fixes the position of the single source at
\beq
z_0= \frac{z_1+z_2}{2} .
\eeq
Then we require $M_0=\bar{M}$,
that fixes $m_0$ and $z_2$ as
\begin{align}
m_0 & = 2 m_1 \sqrt{\frac{2m_1-z_1+z_2}{z_2-z_1}} , \\
z_2 & = z_1 + (\sqrt{17}-1) m_1 .
\end{align}
The entropy variation is given by
$\Delta S=S_0-\bar{S}$,
where the value of the parameters found above must be substituted into the expression.
The result, which is a function of $m_1$ and $z_1$, is quite involved, but it can be plotted as in Fig.~\ref{fig:entropy}.
It is clear from the plot that the function is always positive or null, i.e.
\beq
\Delta S \geq 0 ,
\eeq
which verifies the second law of thermodynamics.
This is one of the few cases in which the second law for a binary system can be verified analytically;
it has been done, e.g., for the Majumdar--Papapetrou solution~\cite{Astorino:2019ljy}.

\section{Conclusions}

In this Letter we built the first analytical solution describing two black holes at equilibrium in General Relativity.
Its feasibility relies on two novel fundamental features for this kind of model:
$(i)$ the spacetime is completely regular outside the black hole horizons and ($ii$) the metric, thanks to its multipolar expansion, can be naturally embedded in phenomenological settings, such as the center of galaxies.
Thus the presented solution can be considered of some realistic astrophysical interest, or even as a reference for Numerical Relativity.
In fact it is possible to transform perturbatively our binary model to a rotating frame (relative to the center of mass of the sources)~\cite{Ni:1978zz}.
Obviously, it would be necessary a solution generating technique that deals with a metric depending on three coordinates to exactly model the gravitational coalescence of two sources, which is an intrinsically dynamical phenomenon; unfortunately such a method is not available at the moment.
%Of course, to model a strong interaction of gravitational sources, which is an intrinsically dynamical phenomenon, it would be necessary a solution generation technique which deals with three coordinates dependence, but unfortunately it is not available at the moment.
However the system described in this work could be at least interpreted, in the above co-rotating frame and at first approximation, as a stationary phase of a binary black hole interaction during a merging process.

Generalisation to an infinite array of sources can be achieved in a straightforward manner:
to this end, the complete multipolar expansion of the external gravitational field will be necessary to obtain a regular solution.
Also the inclusion of electromagnetic charge, NUT and angular momentum represents a natural generalisation of the spacetime studied here~\cite{Astorino:2021boj}. \\

\textbf{Aknowledgments.}
The authors are grateful to Roberto Cotesta and Matteo Macchini for useful discussions, and Roberto Emparan for enlightening correspondence.
They also thank Matteo Broccoli and Sergio Cacciatori for critical readings of the draft of this paper.
This work was supported in part by Conicyt--Beca Chile n$^\textrm{o}$ 74200076, in part by MIUR-PRIN contract 2017CC72MK003 and also by INFN.

%\bibliographystyle{unsrt} %references ordered by citations in the main text
%\bibliographystyle{unsrtnat} %references ordered by citations in the main text + clickable links but ugly links!
%\bibliographystyle{apsrev4-1} %references ordered by citations in the main text + clickable links but no titles!

%SOLUTION
%If I put "longbibliography" global option and no "\bibliographystyle", then I have the beauty of apsrev4-1 + the titles!

\bibliography{Ref.bib}

\end{document}